\documentclass[]{article}
\input epsf
\catcode`\@=11
\long\def\@makefntext#1{
\protect\noindent \hbox to 3.2pt {\hskip-.9pt  

$^{{\eightrm\@thefnmark}}$\hfil}#1\hfill}		%CAN BE USED 

\def\@makefnmark{\hbox to 0pt{$^{\@thefnmark}$\hss}}	%ORIGINAL 

\def\ps@myheadings{\let\@mkboth\@gobbletwo
\def\@oddhead{\hbox{}
\rightmark\hfil\eightrm\thepage}   

\def\@oddfoot{}\def\@evenhead{\eightrm\thepage\hfil
\leftmark\hbox{}}\def\@evenfoot{}
\def\sectionmark##1{}\def\subsectionmark##1{}}

\newcounter{sectionc}\newcounter{subsectionc}\newcounter{subsubsectionc}
\renewcommand{\section}[1] {\vspace{12pt}\addtocounter{sectionc}{1} 

\setcounter{subsectionc}{0}\setcounter{subsubsectionc}{0}\noindent 

	{\tenbf\thesectionc. #1}\par\vspace{5pt}}
\renewcommand{\subsection}[1] {\vspace{12pt}\addtocounter{subsectionc}{1} 

	\setcounter{subsubsectionc}{0}\noindent 

	{\bf\thesectionc.\thesubsectionc. {\kern1pt \bfit #1}}\par\vspace{5pt}}
\renewcommand{\subsubsection}[1] {\vspace{12pt}\addtocounter{subsubsectionc}{1}
	\noindent{\tenrm\thesectionc.\thesubsectionc.\thesubsubsectionc.
	{\kern1pt \tenit #1}}\par\vspace{5pt}}
\newcommand{\nonumsection}[1] {\vspace{12pt}\noindent{\tenbf #1}
	\par\vspace{5pt}}

\newcounter{appendixc}
\newcounter{subappendixc}[appendixc]
\newcounter{subsubappendixc}[subappendixc]
\renewcommand{\thesubappendixc}{\Alph{appendixc}.\arabic{subappendixc}}
\renewcommand{\thesubsubappendixc}
	{\Alph{appendixc}.\arabic{subappendixc}.\arabic{subsubappendixc}}

\renewcommand{\appendix}[1] {\vspace{12pt}
        \refstepcounter{appendixc}
        \setcounter{figure}{0}
        \setcounter{table}{0}
        \setcounter{lemma}{0}
        \setcounter{theorem}{0}
        \setcounter{corollary}{0}
        \setcounter{definition}{0}
        \setcounter{equation}{0}
        \renewcommand{\thefigure}{\Alph{appendixc}.\arabic{figure}}
        \renewcommand{\thetable}{\Alph{appendixc}.\arabic{table}}
        \renewcommand{\theappendixc}{\Alph{appendixc}}
        \renewcommand{\thelemma}{\Alph{appendixc}.\arabic{lemma}}
        \renewcommand{\thetheorem}{\Alph{appendixc}.\arabic{theorem}}
        \renewcommand{\thedefinition}{\Alph{appendixc}.\arabic{definition}}
        \renewcommand{\thecorollary}{\Alph{appendixc}.\arabic{corollary}}
        \renewcommand{\theequation}{\Alph{appendixc}.\arabic{equation}}
        \noindent{\tenbf Appendix \theappendixc #1}\par\vspace{5pt}}
\newcommand{\subappendix}[1] {\vspace{12pt}
        \refstepcounter{subappendixc}
        \noindent{\bf Appendix \thesubappendixc. {\kern1pt \bfit #1}}
	\par\vspace{5pt}}
\newcommand{\subsubappendix}[1] {\vspace{12pt}
        \refstepcounter{subsubappendixc}
        \noindent{\rm Appendix \thesubsubappendixc. {\kern1pt \tenit #1}}
	\par\vspace{5pt}}

\newcommand{\smalllineskip}{\baselineskip=10pt}

\def\eightcirc{
\begin{picture}(0,0)
\put(4.4,1.8){\circle{6.5}}
\end{picture}}
\def\eightcopyright{\eightcirc\kern2.7pt\hbox{\eightrm c}}

\newcommand{\copyrightheading}[1]
	{\vspace*{-2.5cm}\smalllineskip{\flushleft
	{\footnotesize PAR--LPTHE 96--31}\\
	{\footnotesize hep-th/9609049}\\
	 }}

\def\abstracts#1#2#3{{
	\centering{\begin{minipage}{4.5in}\baselineskip=10pt\footnotesize
	\parindent=0pt #1\par 

	\parindent=15pt #2\par
	\parindent=15pt #3
	\end{minipage}}\par}}

\renewenvironment{thebibliography}[1]
	{\frenchspacing
	 \ninerm\baselineskip=11pt
	 \begin{list}{\arabic{enumi}.}
	{\usecounter{enumi}\setlength{\parsep}{0pt}
	 \setlength{\leftmargin 12.7pt}{\rightmargin 0pt} %FOR 1--9 ITEMS
	 \setlength{\itemsep}{0pt} \settowidth
	{\labelwidth}{#1.}\sloppy}}{\end{list}}

\newcounter{itemlistc}
\newcounter{romanlistc}
\newcounter{alphlistc}
\newcounter{arabiclistc}

\newcommand{\fcaption}[1]{
        \refstepcounter{figure}
        \setbox\@tempboxa = \hbox{\footnotesize Fig.~\thefigure. #1}
        \ifdim \wd\@tempboxa > 5in
           {\begin{center}
        \parbox{5in}{\footnotesize\smalllineskip Fig.~\thefigure. #1}
            \end{center}}
        \else
             {\begin{center}
             {\footnotesize Fig.~\thefigure. #1}
              \end{center}}
        \fi}

\newcommand{\tcaption}[1]{
        \refstepcounter{table}
        \setbox\@tempboxa = \hbox{\footnotesize Table~\thetable. #1}
        \ifdim \wd\@tempboxa > 5in
           {\begin{center}
        \parbox{5in}{\footnotesize\smalllineskip Table~\thetable. #1}
            \end{center}}
        \else
             {\begin{center}
             {\footnotesize Table~\thetable. #1}
              \end{center}}
        \fi}

\def\pmb#1{\setbox0=\hbox{#1}
	\kern-.025em\copy0\kern-\wd0
	\kern.05em\copy0\kern-\wd0
	\kern-.025em\raise.0433em\box0}

\def\fnt#1#2{\footnotetext{\kern-.3em
	{$^{\mbox{\scriptsize #1}}$}{#2}}}

\font\tenrm=cmr10
\font\tenit=cmti10 

\font\tenbf=cmbx10
\font\bfit=cmbxti10 at 10pt
\font\ninerm=cmr9

\font\eightrm=cmr8

\def\qed{\hbox{${\vcenter{\vbox{			%HOLLOW SQUARE
   \hrule height 0.4pt\hbox{\vrule width 0.4pt height 6pt
   \kern5pt\vrule width 0.4pt}\hrule height 0.4pt}}}$}}

\def\be{\begin{equation}}
\def\ee{\end{equation}}
\def\bea{\begin{eqnarray}}
\def\eea{\end{eqnarray}}

\begin{document}

\title{{\large \bf Charmonium Suppression in Nuclear Collisions}}
\author{{\small Sean Gavin}
\thanks{This work was supported in part by US-DOE contracts 
        DE-FG02-93ER40764 and DE-AC02-76CH00016.}\\ 
{\small \it Physics Department, Brookhaven National Laboratory, Upton, NY}
\thanks{Mailing address.}\\
{\small \it and}\\ 
{\small \it Department of Physics, Columbia University, New York, NY.}}

\date{\small September 1996} 
\maketitle 
\abstracts{Measurements of $\psi$ and $\psi^\prime$ production 
from experiment NA50 at the CERN SPS are compared to calculations 
based on a hadronic model of charmonium suppression developed
previously.  Data on centrality dependence and total cross sections
are in good accord with these predictions.  Uncertainties in theoretical 
quantities such as NA50's $L$ parameter are discussed.
}{}{}

\section{Introduction}
\noindent
Has the quark gluon plasma been discovered at the CERN SPS?
Experiment NA50 has reported an abrupt decrease in
$\psi$ production in Pb+Pb collisions at 158 GeV per nucleon
\cite{na50}.  Specifically, the collaboration presented a striking
`threshold effect' in the $\psi$--to--continuum ratio by plotting it
as a function of a calculated quantity, the mean path length of the
$\psi$ through the nuclear medium, $L$, as shown in fig.~1a.  This apparent
threshold has sparked considerable excitement as it may signal 
deconfinement in the heavy Pb+Pb system \cite{bo}.
\begin{figure}
\vskip -1.0in
\epsfxsize=4.5in
\leftline{\epsffile{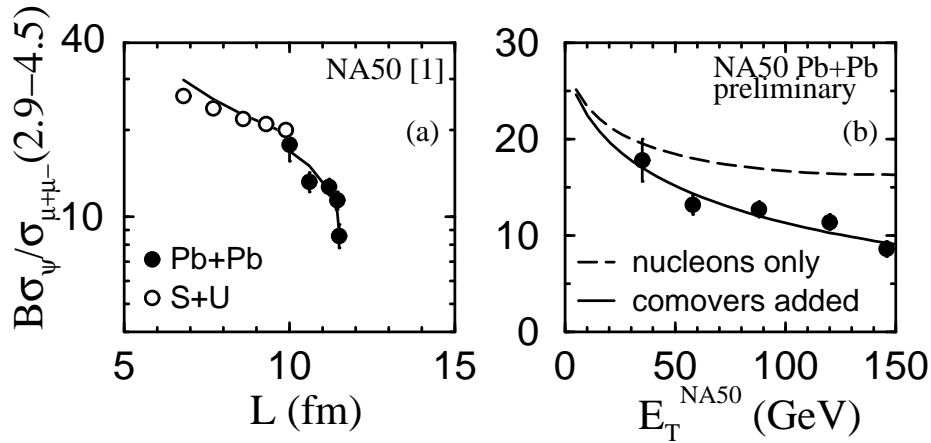}}
\vskip -2.4in
\caption[]{(a) The NA50 \cite{na50} comparison of $\psi$ production in
Pb+Pb and S+U collisions as a function of the average path length $L$,
see eq.\ (3).  $B$ is the $\psi\rightarrow \mu^+\mu^-$ branching
ratio. (b) Transverse energy dependence of Pb+Pb data.  Curves in (a)
and (b) are computed using eqs.\ (4--6).}
\end{figure}

In this talk I report on work with Ramona Vogt in ref.~\cite{gv2}
comparing Pb results to predictions \cite{gv,gstv} using a hadronic
model of charmonium suppression.  We first demonstrate
that the behavior in the NA50 plot, fig.~1a, is not a threshold effect
but, rather, reflects the approach to the geometrical limit of $L$ as
the collisions become increasingly central.  When plotted as a
function of the {\it measured} neutral transverse energy $E_{T}$ as in
fig.~1b, the data varies smoothly as in S+U measurements in fig.~3b
below \cite{na50,na38,na38c,na38d,na38e}.  The difference between S+U
and Pb+Pb data lies strictly in the relative magnitude.  To assess
this magnitude, we compare $\psi$ and $\psi^\prime$ data to
expectations based on the hadronic comover model \cite{gv,gstv}.  The
curves in fig.~1 represent our calculations using parameters fixed
earlier in Ref.\ \cite{gstv}.  Our result is essentially the same as
the Pb+Pb prediction in \cite{gv}.

Our primary intention is to demonstrate that there is no evidence for
a strong discontinuity between $p$A, S+U and Pb+Pb data.  However, to
quote Maurice Goldhaber, ``$\ldots$ absence of evidence is {\it not}
evidence of absence.''  Our secondary goal is to show that our model
predictions agree with the new Pb+Pb data.  The consistency of these
predictions is evident from the agreement of our old $p$A and S+U
calculations with more recent NA38 and NA51 data.  Nevertheless, the
significance of this result must be weighted by the fact that all
$p$A and AB data are preliminary and at different beam energies.

In this work, we do not attempt to show that our comover
interpretation of the data is unambiguous -- this is certainly
impossible at present.

\section{Nucleons and Comovers}
The hadronic contribution to charmonium suppression arises from scattering of
the nascent $\psi$ with produced particles -- the comovers -- and
nucleons \cite{gv,gstv}.  To determine the suppression from nucleon
absorption of the $\psi$, we calculate the probability that a
$c{\overline c}$ pair produced at a point $(b, z)$ in a nucleus
survives scattering with nucleons to form a $\psi$.  The standard
\cite{gstv,gh} result is
\begin{equation}
S_{A} = {\rm exp}\{-\int_z^\infty\! dz\, \rho_{A}(b, z) \sigma_{\psi N}\}
\end{equation}
where $\rho_{A}$ is the nuclear density, $b$ the impact parameter and
$\sigma_{\psi N}$ the absorption cross section for $\psi$--nucleon
interactions.  One can estimate $S_{A}\sim \exp\{-
\sigma_{\psi N} \rho_0 L_{A}\}$, where $L_{A}$ is the path length
traversed by the $c\overline{c}$ pair.

Suppression can also be caused by scattering with mesons that happen
to travel along with the $c\overline{c}$ pair (see refs.\ in
\cite{gv}).  The density of such comovers scales roughly as $E_{T}$.
The corresponding survival probability is
\begin{equation}
S_{\rm co} = {\rm exp}\{- \int\! d\tau n\, 
\sigma_{\rm co} v_{\rm rel}\},
\end{equation}
where $n$ is the comover density and $\tau$ is the time in the $\psi$
rest frame.  We write $S_{\rm co}\sim {\rm exp}\{-\beta E_{T}\}$,
where $\beta$ depends on the scattering frequency,
the formation time of the comovers and the transverse size of the
central region, $R_{T}$, {\it cf.} eq.\ (8).

To understand the saturation of the Pb data with $L$ in fig.~1a, we apply
the schematic approximation of Ref.~\cite{gh} for the moment to write
\begin{equation}
{{\sigma^{AB}_\psi(E_{T})}\over{\sigma^{AB}_{\mu^+\mu^-}(E_{T})}}
\propto \langle S_{A}S_{B}S_{\rm co}\rangle
\sim 
{\rm e}^{-\sigma_{\psi N}\rho_{0}L}{\rm e}^{-\beta E_{T}},
\end{equation}
where the brackets imply an average over the collision geometry for
fixed $E_{T}$ and $\sigma(E_T) \equiv d\sigma/dE_T$.  The path length
$L\equiv \langle L_{A}+L_{B}\rangle$ and transverse size $R_T$ depend
on the collision geometry.  The path length grows with $E_{T}$,
asymptotically approaching the geometric limit $R_A + R_B$.  Explicit
calculations show that nucleon absorption begins to {\it saturate} for
$b < R_A$, where $R_A$ is the smaller of the two nuclei, see fig.~4
below.  On the other hand, $E_{T}$ continues to
grow for $b < R_A$ due, {\it e.g.}, to fluctuations in the number of
$NN$ collisions.  Equation (2) falls exponentially in this regime
because $\beta$, like $L$, saturates.

In fig.~1b, we compare the Pb data to calculations of the
$\psi$--to--continuum ratio that incorporate nucleon and comover
scattering.  The contribution due to nucleon absorption indeed levels
off for small values of $b$, as expected from eq.\ (3).  Comover
scattering accounts for the remaining suppression.
 
These results are {\it predictions} obtained using the computer code
of Ref.~\cite{gv} with parameters determined in Ref.~\cite{gstv}.
However, to confront the present NA50 analysis \cite{na50}, we 
account for changes in the experimental coverege as follows:
\begin{itemize}
\item Calculate the continuum dimuon yield in the new mass range $2.9
< M < 4.5$~GeV.  
\item Adjust the $E_T$ scale to the pseudorapidity
acceptance of the NA50 calorimeter, $1.1 < \eta < 2.3$.
\end{itemize} 
The agreement in fig.~1 depends on these updates.

\section{$J/\psi$ Suppression}
We now review the details of our calculations, highlighting the
adjustments as we go.  For collisions at a fixed $b$, the
$\psi$--production cross section is
\begin{equation}
\sigma_\psi^{AB}(b)
=
\sigma^{NN}_{\psi}\!\int\! d^2s dz dz^\prime\,\rho_A(s,z)
\rho_B(b-s,z^\prime)\, S,
\end{equation}
where $S\equiv S_AS_BS_{\rm co}$ is the product of the survival
probabilities in the projectile $A$, target $B$ and comover matter.
The continuum cross section is
\begin{equation}
\sigma_{\mu^{+}\mu^{-}}^{AB}(b) = 
\sigma^{NN}_{\mu^+\mu^-}\!\int\! d^2s dz dz^\prime\,\rho_A(s,z)
\rho_B(b-s,z^\prime).
\end{equation}
The magnitude of (4,5) and their ratio are fixed by the elementary
cross sections $\sigma^{NN}_{\psi}$ and
$\sigma^{NN}_{\mu^{+}\mu^{-}}$.  We calculate $\sigma^{NN}_{\psi}$
using the phenomenologically--successful color evaporation model
\cite{hpc-psi}.  The continuum in the mass range used by NA50, $2.9 <
M < 4.5$~GeV, is described by the Drell--Yan process.  To confront
NA50 and NA38 data in the appropriate kinematic regime, we compute
these cross sections at leading order following \cite{hpc-psi,hpc-dy}
using GRV LO parton distributions with a charm $K$--factor $K_c= 2.7$
and a color evaporation coefficient $F_\psi =2.54\%$ and a Drell--Yan
$K$--factor $K_{DY}=2.4$.  Observe that these choices were fixed by
fitting $pp$ data at all available energies \cite{hpc-psi}.  Computing
$\sigma^{NN}_{\mu^{+}\mu^{-}}$ for $2.9<M<4.5$~GeV corresponds to the
first update.

To obtain $E_T$ dependent cross sections from eqs.\ (4) and (5), we
write
\begin{equation}
\sigma^{AB}(E_{T}) =
\int\! d^2b\, P(E_T,b) \sigma^{AB}(b).
\end{equation}
The probability $P(E_T,b)$ that a collision at impact parameter $b$
produces transverse energy $E_T$ is related to the minimum--bias
distribution by
\begin{equation}
\sigma_{\rm min}(E_{T}) = \int\! d^{2}b\; P(E_{T}, b).
\end{equation}
We parametrize $P(E_{T}, b) = C\exp\{- (E_{T}- {\overline
E}_{T})^2/2\Delta\}$, where ${\overline E}_{T}(b) = \epsilon {\cal
N}(b)$, $\Delta(b) = \omega \epsilon {\overline E}_{T}(b)$,
$C(b)=(2\pi\Delta(b))^{-1}$ and ${\cal N}(b)$ is the number of
participants (see, {\it e.g.}, Ref.~\cite{gv}).
We take $\epsilon$ and $\omega$ to be phenomenological
calorimeter--dependent constants.

We compare the minimum bias distributions for total hadronic $E_T$
calculated using eq.\ (7) for $\epsilon = 1.3$~GeV and $\omega = 2.0$
to NA35 S+S and NA49 Pb+Pb data \cite{na49}.  The agreement in fig.~2a
builds our confidence that eq.\ (7) applies to the heavy Pb+Pb system.
\begin{figure}
\vskip -1.5in
\epsfxsize=4.0in
\centerline{\epsffile{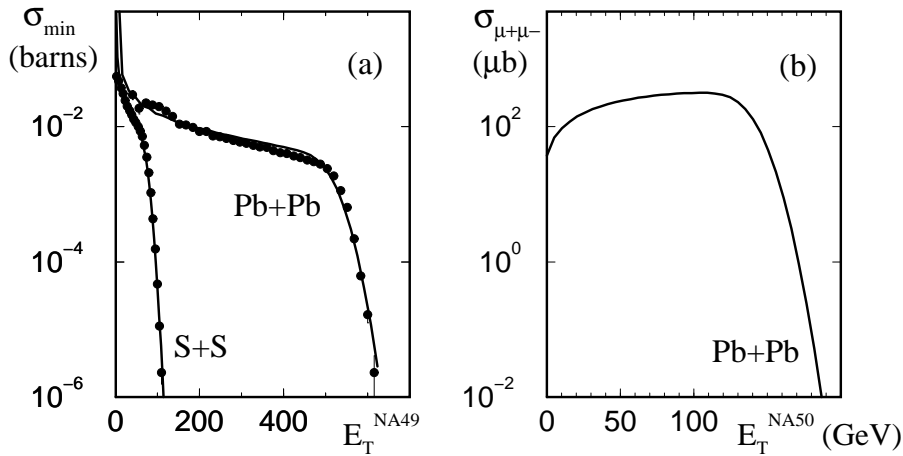}}
\vskip -1.0in
\caption{Transverse energy distributions from eq.\ (7).
The S--Pb comparison (a) employs the same parameters.}
\end{figure}
Figure 2b shows the distribution of neutral transverse energy
calculated using eqs.\ (5) and (6) to simulate the NA50 dimuon
trigger. We take $\epsilon = 0.35$~GeV, $\omega = 3.2$, and
$\sigma^{NN}_{\mu^+\mu^-}\approx 37.2$~pb as appropriate for the
dimuon--mass range $2.9 < M < 4.5$~GeV.  The $E_T$ distribution for
S+U~$\rightarrow \mu^+\mu^- + X$ from NA38 was described \cite{gstv}
using $\epsilon = 0.64$~GeV and $\omega = 3.2$ -- the change in
$\epsilon$ corresponds roughly to the shift in particle production
when the pseudorapidity coverage is changed from $1.7 < \eta < 4.1$
(NA38) to $1.1 < \eta < 2.3$ (NA50).  Taking $\epsilon = 0.35$~GeV for
the NA50 acceptance is the second update listed earlier.
We now apply eqs.\ (1,2,4) and (5) to charmonium suppression in Pb+Pb
collisions.  To determine nucleon absorption, we used $p$A data to fix
$\sigma_{\psi N}\approx 4.8$~mb in Ref.~\cite{gstv}. This choice is in
accord with the latest NA38 and NA51 $pA$ data, see fig.~3a.
To specify comover scattering \cite{gstv}, we assumed that the
dominant contribution to $\psi$ dissociation comes from exothermic
hadronic reactions such as $\rho + \psi \rightarrow D+ \overline{D}$.
We further took the comovers to evolve from a formation time
$\tau_{0}\sim 2$~fm to a freezeout time $\tau_{F}\sim R_{T}/v_{\rm
rel}$ following Bjorken scaling, where $v_{\rm rel}\sim 0.6$ is
roughly the average $\psi-\rho$ relative velocity.  The
survival probability, eq.\ (2), is then
\begin{equation}
S_{\rm co} = \exp\{ - \sigma_{\rm co}v_{\rm rel}n_{0}\tau_{0} 
\ln(R_{T}/v_{\rm rel}\tau_{0})\}
\end{equation}
where $\sigma_{\rm co} \approx 2\sigma_{\psi N}/3$, $R_{T}\approx
R_{A}$ and $n_{0}$ is the initial density of sufficiently massive
$\rho, \omega$ and $\eta$ mesons.  To account for the variation of
density with $E_{T}$, we take $n_{0} = {\overline
n}_{0}E_{T}/{\overline E}_{T}(0)$ \cite{gv}. A value $\overline{n}_{0}
= 0.8$~fm$^{-3}$ was chosen to fit the central S+U datum.  Since we
fix the density in central collisions, this simple {\it ansatz} for
$S_{\rm co}$ may be inaccurate for peripheral collisions.  [Densities
$\sim 1$~fm$^{-3}$ typically arise in hadronic models of ion
collisions, e.g., refs.~\cite{cascade}.  The internal consistency of
hadronic models at such densities demands further study.]

We expect the comover contribution to the suppression to increase in
Pb+Pb relative to S+U for central collisions because both the
initial density and lifetime of the system can increase.  To be
conservative, we assumed that Pb and S beams achieve the same mean initial
density.  Even so, the lifetime of the system essentially doubles in
Pb+Pb because $R_T \sim R_{A}$ increases to 6.6~fm from 3.6~fm in S+U.
The increase in the comover contribution evident in comparing figs.~1b
and 3b is described by the seemingly innocuous logarithm in eq.\ (8),
which increases by $\approx 60\%$  in the larger Pb system.
\begin{figure}
\vskip -2.8in
\epsfxsize=4.5in
\rightline{\epsffile{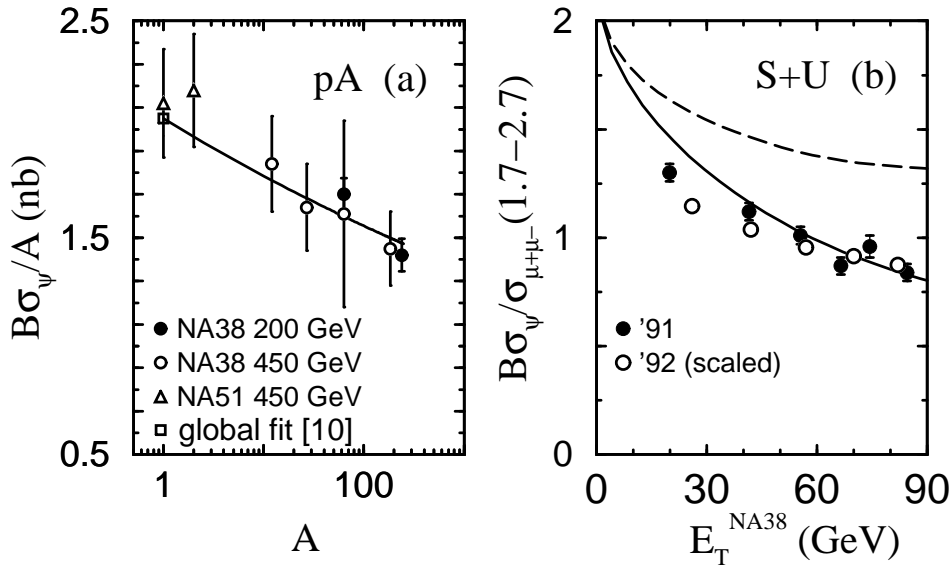}}
\vskip -0.5in
\caption[]{(a) $p$A cross sections \cite{na50} in the NA50 acceptance
and (b) S+U ratios from '91 \cite{na38c} and '92 \cite{na50}
runs. The '92 data are scaled to the '91 continuum.  The dashed line
indicates the suppression from nucleons alone.  The $pp$ cross section
in (a) is constrained by the global fit to $pp$ data in
ref.~\cite{hpc-psi}.}
\end{figure}

In Ref.~\cite{gstv}, we pointed out that comovers were necessary to
explain S+U data from the NA38 1991 run \cite{na38}.  Data just
released \cite{na50} from their 1992 run support this conclusion.  The
'91 $\psi$ data were presented as a ratio to the dimuon continuum in
the low mass range $1.7 < M < 2.7$~GeV, where charm decays are an
important source of dileptons.  On the other hand, the '92 $\psi$ data
\cite{na50,na38e} are given as ratios to the Drell--Yan cross section
in the range $1.5< M < 5.0$~GeV.  That cross section is extracted from
the continuum by fixing the $K$--factor in the high mass region
\cite{na38f}.  To compare our result from Ref.~\cite{gstv} to these
data, we scale the '92 data by an empirical factor.  This factor is
$\approx 10\%$ larger than our calculated factor
$\sigma^{NN}_{DY}(92)/\sigma^{NN}_{\rm cont.}(91) \approx 0.4$; these
values agree within the NA38 systematic errors.  [NA50 similarly
scaled the '92 data to the high--mass continuum to produce fig.~1a.]
Because our fit is driven by the highest $E_T$ datum, we see from
fig.~3b that a fit to the '92 data would not appreciably change our
result.  Note that a uniform decrease of the ratio would increase the
comover contribution needed to explain S+U collisions.

NA50 and NA38 have also measured the total $\psi$--production cross
section in Pb+Pb \cite{na50} and S+U reactions \cite{na38c}.  To
compare to that data, we integrate eqs.\ (4, 6) to obtain the total
$(\sigma/AB)_{\psi} = 0.95$~nb in S+U at 200~GeV and 0.54~nb for Pb+Pb
at 158~GeV in the NA50 spectrometer acceptance, $0.4 > x_{F}> 0$ and
$-0.5 < \cos\theta < 0.5$ (to correct to the full angular range and $1
> x_{F} > 0$, multiply these cross sections by $\approx 2.07$).  The
experimental results in this range are $1.03 \pm 0.04 \pm 0.10$~nb for
S+U collisions \cite{na38} and $0.44 \pm 0.005 \pm 0.032$ nb for Pb+Pb
reactions \cite{na50}.  Interestingly, in the Pb system we find a
Drell--Yan cross section $(\sigma/AB)_{{}_{DY}} = 37.2$~pb while NA50
finds $(\sigma/AB)_{{}_{DY}} = 32.8\pm 0.9\pm 2.3$~pb.  Both the
$\psi$ and Drell--Yan cross sections in Pb+Pb collisions are somewhat
above the data, suggesting that the calculated rates at the $NN$ level
may be $\sim 20-30\%$ too large at 158~GeV.  This discrepancy is
within ambiguities in current $pp$ data near that low energy
\cite{hpc-psi}.  Moreover, nuclear effects on the parton densities
omitted in eqs.\ (4,5) can affect the total S and Pb cross sections at
this level.

We remark that if one were to neglect comovers and take $\sigma_{\psi
N} = 6.2$~mb, one would find $(\sigma/AB)_{\psi} = 1.03$~nb in S+U at
200~GeV and 0.62~nb for Pb+Pb at 158~GeV.  The agreement with S+U data
is possible because comovers only contribute to the total cross
section at the $\sim 18\%$ level in the light system.  This is
expected, since the impact--parameter integrated cross section is
dominated by large $b$ and the distinction between central and
peripheral interactions is more striking for the asymmetric S+U
system.  As in Ref.~\cite{gstv}, the need for comovers is evident for
the $E_{T}$--dependent ratios, where central collisions are singled
out.

\section{Saturation and the Definition of $L$}
To see why saturation occurs in Pb+Pb collisions but not in S+U, we
compare the NA50 $L(E_T)$ \cite{na50} to the average impact parameter
$\langle b\rangle (E_T)$ in fig.~4.  To best understand fig.~1a, we
show the values of $L(E_T)$ computed by NA50 for this figure.  We use
our model to compute $\langle b\rangle = \langle b
T_{AB}\rangle/\langle T_{AB}\rangle$, where $\langle f(b)\rangle
\equiv \int\!d^2b\; P(E_T,b)f(b)$ and $T_{AB} =
\int\!d^{2}sdzdz^\prime \rho_{A}(s,z)\rho_{B}(b-s,z^\prime)$. [Note
that NA50 reports similar values of $\langle b\rangle (E_T)$
\cite{na50}.] In the $E_T$ range covered by the S experiments, we see
that $\langle b\rangle$ is near $\sim R_{\rm S} = 3.6$~fm or larger.
In this range, increasing $b$ dramatically reduces the collision
volume and, consequently, $L$.  In contrast, in Pb+Pb collisions
$\langle b\rangle \ll R_{\rm Pb} =$~6.6~fm for all but the lowest
$E_T$ bin, so that $L$ does not vary appreciably.
\begin{figure}
\vskip -2.8in
\epsfxsize=4.5in
\rightline{\epsffile{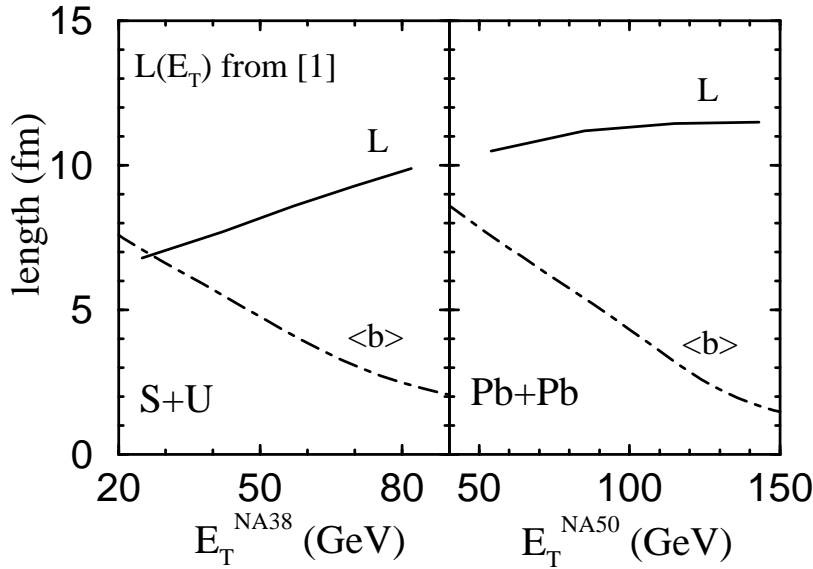}}
\vskip -0.5in
\caption[]{$E_T$ dependence of $L$ (solid) used by NA50 \cite{na50}
(see fig.~1a) and the average impact parameter $\langle b\rangle$
(dot--dashed).  The solid line covers the measured $E_T$ range.}
\end{figure}
\begin{figure}
\vskip -2.8in
\epsfxsize=4.5in
\rightline{\epsffile{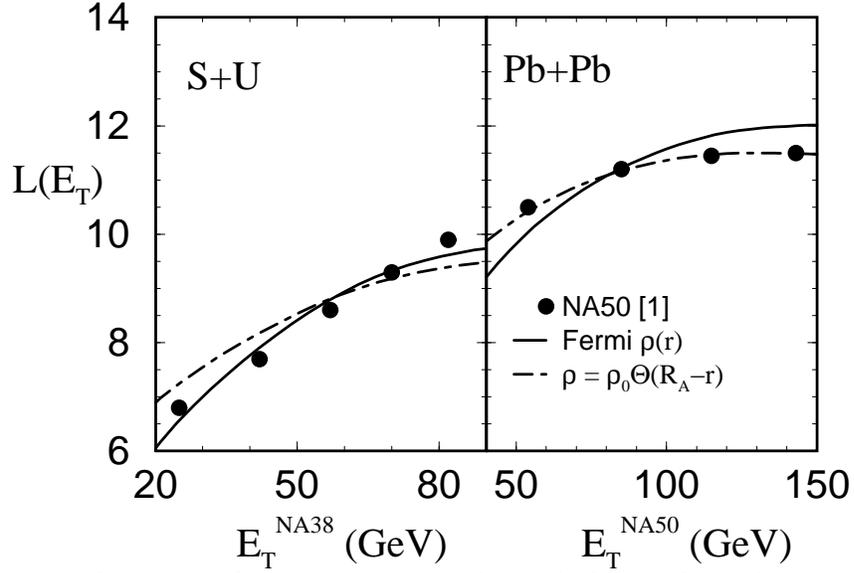}}
\vskip -0.5in
\caption[]{NA50 $L(E_T)$ [1] (points) compared to calculations
for realistic nuclear densities (solid), as used here, and for
a sharp--surface approximation (dot-dashed).}
\end{figure}
\begin{figure}
\vskip -2.0in
\epsfxsize=4.5in
\centerline{\epsffile{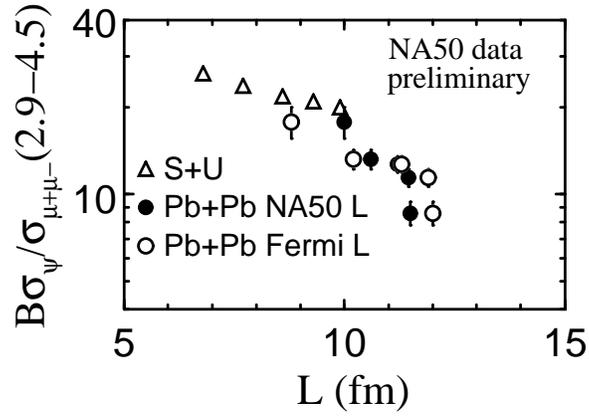}}
\vskip -2.0in
\caption{NA50 data replotted with a realistic $L(E_T)$ from (9).}
\end{figure}
To understand the sensitivity of fig.~1a to the definition of the path
length, we now estimate $L(E_T)$ \cite{gv3}.  We identify (3) with the
exact expression formed from the ratio of (4) and (5).  Expanding in
$\sigma_{\psi N}$ and neglecting comovers, we find:
\begin{equation}
L(E_T) = \{2\rho_0\langle T_{AB}\rangle\}^{-1}
\left\langle\int\! d^2s\; [T_A(s)]^2T_B(b-s) + 
[T_B(b-s)]^2T_A(s)\right\rangle,
\end{equation}
where $T_A(s) = \int \rho_A(s,z) dz$.
In fig.~5 we compare the NA50 $L(E_T)$ to the path length calculated
using two assumptions for the nuclear density profile: our realistic
three--parameter Fermi distribution and the sharp--surface
approximation $\rho = \rho_0\Theta(R_A -r)$.  NA38~\cite{borhani}
obtained $L$ for S+U using the empirical prescription of
ref.~\cite{gh}, while NA50 calculated $L$ assuming the sharp-surface
approximation~\cite{claudie}.  Indeed, we see that the NA50 Pb+Pb
values agree with our sharp--surface result, while the NA38 S+U values
are nearer to the realistic distribution.

To see how the value of the path length can affect the appearance of
fig.~1a, we replot in fig.~6 the NA50 data using $L(E_T)$ from (9)
with the realistic density.  We learn that the appearance of fig.~1a
is very sensitive to the definition of $L$.  Furthermore, with a
realistic $L$, one no longer gets the impression given by the NA50
figure \cite{na50} of Pb+Pb data ``departing from a universal curve.''
Nevertheless, the saturation phenomena evident in fig.~1a does not
vanish.  Saturation is a real effect of geometry.

\section{$\psi^\prime$ Suppression} 
\begin{figure}
\vskip -3.2in
\epsfxsize=4.5in
\rightline{\epsffile{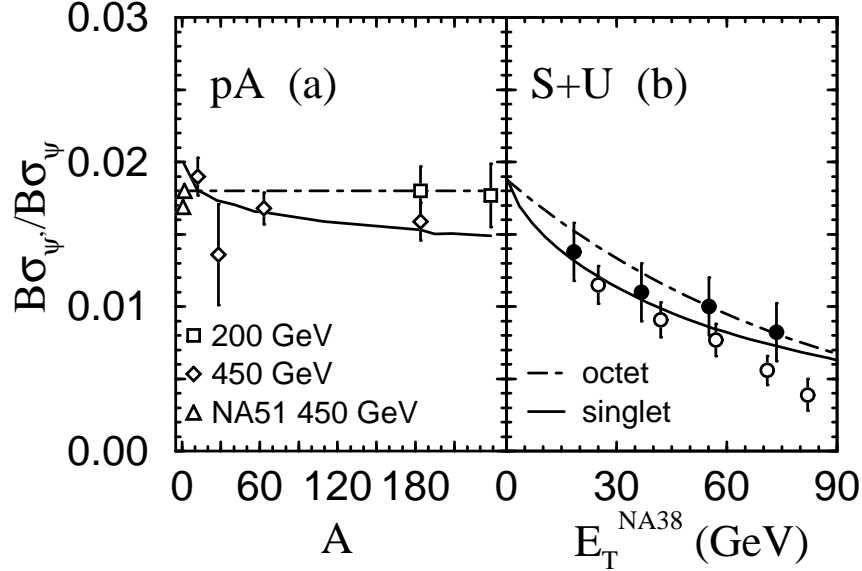}}
\vskip -0.3in
\caption[]{Comover suppression of $\psi^\prime$ compared to (a) NA38 and
NA51 $p$A data \cite{na50,na38e} and (b) NA38 S+U data \cite{na38d}
(filled points) and preliminary data \cite{na50}.}
\end{figure}
\begin{figure}
\vskip -2.0in
\epsfxsize=4.5in
\centerline{\epsffile{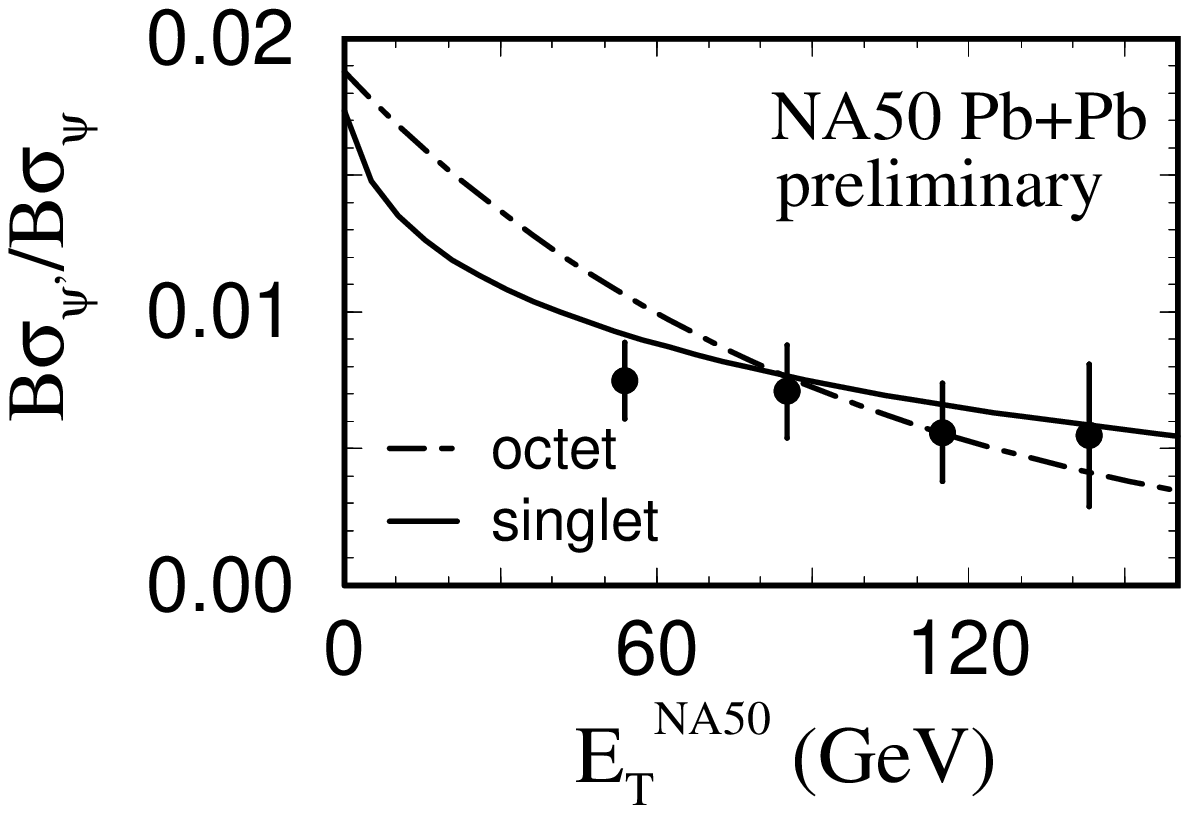}}
\vskip -2.0in
\caption{Comover suppression in Pb+Pb~$\rightarrow \psi^\prime +X$.}
\end{figure}
To apply eqs.\ (4-6) to calculate the $\psi^{\prime}$--to--$\psi$
ratio as a function of $E_{T}$, we must specify
$\sigma_{\psi^{\prime}}^{NN}$, $\sigma_{\psi^{\prime} N}$, and
$\sigma_{\psi^{\prime} {\rm co}}$.  Following Ref.~\cite{hpc-psi}, we
use $pp$ data to fix $B\sigma_{\psi^{\prime}}^{NN}/B\sigma_{\psi}^{NN}
= 0.02$ (this determines $F_{\psi^\prime}$).  The value of
$\sigma_{\psi^{\prime} N}$ depends on whether the nascent
$\psi^{\prime}$ is a color singlet hadron or color octet
$c\overline{c}$ as it traverses the nucleus. In the singlet case, one
expects the absorption cross sections to scale with the square of the
charmonium radius.  Taking this {\it ansatz} and assuming that the
$\psi^\prime$ forms directly while radiative $\chi$ decays account for
40\% of $\psi$ production, one expects $\sigma_{\psi'}\sim
2.1\sigma_{\psi}$ for interactions with either nucleons or comovers
\cite{gstv}.  For the octet case, we take $\sigma_{\psi^{\prime} N}
\approx \sigma_{\psi N}$ and fix $\sigma_{\psi^{\prime} {\rm
co}}\approx 12$~mb to fit the S+U data.  In fig.~7a, we show that the
singlet and octet extrapolations describe $p$A data equally well.

Our predictions for Pb+Pb collisions are shown in fig.~8.  In the
octet model, the entire suppression of the $\psi^{\prime}$--to--$\psi$
ratio is due to comover interactions.  In view of the schematic nature
of our approximation to $S_{\rm co}$ in eq.\ (8), we regard the
agreement with data of singlet and octet extrapolations as equivalent.

\section{Summary}
In summary, the Pb data \cite{na50} cannot be described by nucleon
absorption alone.  This is seen in the NA50 plot, fig.~1a, and
confirmed by our results.  The saturation with $L$ but not $E_T$
suggests an additional density--dependent suppression mechanism.
Earlier studies pointed out that additional suppression was already
needed to describe the S+U results \cite{gstv}; recent data
\cite{na50} support that conclusion (see, however, \cite{bo}).
Comover scattering explains the additional suppression.  Nevertheless,
it is unlikely that this explanation is unique.  SPS
inverse--kinematics experiments ($B < A$) and AGS $p$A studies near
the $\psi$ threshold can help pin down model uncertainties.

After the completion of \cite{gv2}, several cascade calculations
\cite{cascade} have essentially confirmed our conclusions.  This
confirmation is important, because such calculations do not employ the
simplifications ({\it e.g.\ } $n_0\propto E_T$) needed to derive (8).
In particular, these models calculate $E_T$ and the comover density
consistently.  Some of these authors took $\sigma_{\psi N} \sim 6$~mb
(instead of $\sim 5$~mb) to fit the NA51 data in fig.~3a somewhat
better.
 
I am grateful to Ramona Vogt for her collaboration in this work.  I
also thank C.~Gerschel and M.~Gonin for discussions of the NA50 data,
and M.~Gyulassy, R.~Pisarski and M.~Tytgat for insightful comments.

\nonumsection{References}

\end{document}